\documentclass{aastex}
\usepackage{spr-astr-addons}
\usepackage{url}

\RequirePackage{color}

\renewcommand{\vec}[1]{\mbox{\boldmath $#1$}}

\newcommand*\Del{\mathrm{\Delta}}                 

\newcommand{\rmd}{{\ \mathrm d} }
\newcommand{\F}{{\ \mathrm F} }

\newcommand{\m}{{\ \mathrm m} }
\newcommand{\s}{{\ \mathrm s} }

\newcommand{\emaila}{wilhelm@mps.mpg.de}
\newcommand{\emailb}{bholadwivedi@gmail.com}

\begin{document}

\title{On the potential energy in an electrostatically bound two-body
system}


\shorttitle{Electrostatic potential energy}
\shortauthors{K. Wilhelm and B.N. Dwivedi}
\author{Klaus Wilhelm}
\affil{Max-Planck-Institut f\"ur Son\-nen\-sy\-stem\-for\-schung
(MPS), 37077~G\"ottingen, Germany \\ \emaila}
\and
\author{Bhola N. Dwivedi}
\affil{Department of Physics, Indian Institute of Technology
(Banaras Hindu University), Varanasi-221005, India \\ \emailb}


\vspace{1cm}

\begin{abstract}
The potential energy problem in an electrostatically bound two-body system
is studied in the framework of a recently proposed impact model of
the electrostatic force
and in analogy to the potential
energy in a gravitationally bound system.
The physical processes are described that result in the variation of the
potential energy as a function of the distance between the charged bodies.
The energy is extracted from distributions of
hypothetical interaction entities modified by the charged bodies.
\end{abstract}


PACS
~~04.20.Cv,      
04.20.Fy,      
04.25.-g,      


\section{Introduction}
\label{s.introd}
In analogy to the gravitational potential treated in
\citet[][Paper~1]{WilDwi},
we can\,--\,according to \citet{LanLif}\,--\,write the Lagrangian of
a closed system consisting of two bodies~A and B in motion with
masses $m_{\rm A}$ and $m_{\rm B}$, respectively,
as
%
\begin{equation}
L = \frac{1}{2}\,(m_{\rm A}\,\vec{V}_{\rm A}^2 + m_{\rm B}\,\vec{V}_{\rm B}^2) -
U(\vec{r}_{\rm A},\vec{r}_{\rm B}) = T - U ~,
\label{eq:Lagrange}
\end{equation}
where $\vec{r}_{\rm A}$, $\vec{r}_{\rm B}$ are the radius vectors of the
bodies and $\vec{V}_{\rm A} = \rmd \vec{r}_{\rm A}/\rmd t$,
$\vec{V}_{\rm B} = \rmd \vec{r}_{\rm B}/\rmd t$ their (non-relativistic)
velocities. The sum~$T$ is the kinetic energy and the function~$U$ here
designates the electrostatic potential energy of the system.
The conservation law of energy can be
derived from the homogeneity of time: The energy $E = T + U$ of
the closed system remains constant during the motion, because $L$ does not
explicitly depend on time.

The external electrostatic potential of a spherically symmetric body~A with
charge $Q$ is
\begin{equation}
\phi_{\rm A}(r) = \frac{Q}{4\,\pi\,\varepsilon_0\,r}  ~,
\label{eq:potential}
\end{equation}
where $r$ is the distance from the centre of the
body \citep[cf., e.g.][]{Jac99}. The electric constant
is $\varepsilon_0 = 8.854\,187\,817... \times 10^{-12}\,\F\,\m^{-1}$
(exact).\footnote{Follows from the definition of
$\mu_0 = 4 \times 10^{–7}\,{\rm H}\,\m^{-1}$, the magnetic constant
(Bureau International des Poids
et Mesures, BIPM, 2006), and $\varepsilon_0 = (\mu_0\,c_0^2)^{-1}$ with
the speed of light in vacuum $c_0 = 299\,792\,458\,\m\,\s^{-1}$
(exact)\,--\,according to the definition of the SI base unit ``metre''.}

Although the introductory statements on the potential energy
are textbook knowledge, \citet{Car98}
wrote: ``\,--\,after all, potential energy is a rather mysterious
quantity to begin with\,--\,''.
This remark motivated us to think about the gravitational potential
energy (Paper~1). Here we will discuss the electrostatic aspects of
the ``mystery''.

\begin{figure}[!t]
\centering
\includegraphics[width=7.6cm]{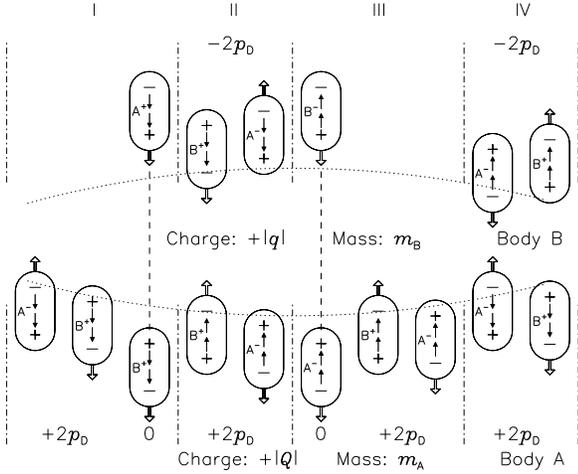}
\caption{\label{fig:plus} The body~A with charge~$+|Q|$ is positioned
in this configuration beneath body~B with charge~$+|q|$ leading to an
electrostatic repulsion of the bodies. This results from
the reversal of dipoles by the charge~$+|Q|$ followed by direct
interactions with the charge~$+|q|$ as defined in Fig.~3 of Paper~2,
where the indirect interactions in columns~I and III are defined as well.
Two reversals are schematically indicated in columns~I and III. The dipoles
arriving in columns~II and IV from below have the same polarity as if they
would be part of the background distribution. The same is true for all
dipoles arriving from above. The net momentum transfer caused by four
interacting dipoles thus is $(8 - 4)\,\vec{p}_{\rm D}$,
i.e. one $\vec{p}_{\rm D}$ per dipole.}
\end{figure}

\begin{figure}[!t]
\centering
\includegraphics[width=7.6cm]{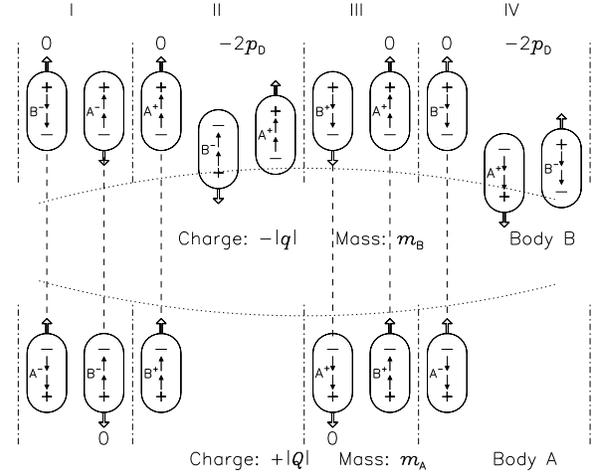}
\caption{\label{fig:minus} The body~A with charge~$+|Q|$ is again positioned
beneath body~B, however, its charge now is~$-|q|$ leading to an
electrostatic attraction of the bodies. The attraction results from
the reversal of dipoles by the charge~$+|Q|$ followed by indirect
interactions with charge~$-|q|$.
Two reversals in columns ~II and IV are schematically indicated. The dipoles
arriving in columns~I and III from below have the same polarity as if they
would be part of the background distribution. The same is true for all
dipoles arriving from above. The net momentum transfer caused by four
interacting dipoles thus is $-4\,\vec{p}_{\rm D}$,
i.e. again one $-\vec{p}_{\rm D}$ per
dipole.}
\end{figure}

In order to have a well-defined configuration for our discussion,
we will assume that body~A has a positive charge~$+|Q|$ and is positioned
beneath body~B with either a charge $+|q|$ in Fig.~\ref{fig:plus} or
$-|q|$ in Fig.~\ref{fig:minus}. Only the processes near the body~B are shown
in detail. Since the electrostatic forces between charged particles~A and B
are typically many orders of magnitude larger than the gravitational forces,
we only take the electrostatic effects into account and neglect the
gravitational interaction.
%
%
%
%

\section{Free particles}
\label{s.free_closed}

The energy~$E_m$ and momentum~$\vec{p}$ of a free particle with mass~$m$
moving with a velocity~$\vec{V}$ relative to an inertial reference system are
related by
%
\begin{equation}
E^2_m - \vec{p}^{\,2}\,c^2_0 = m^2\,c^4_0 ~,
\label{eq:energy}
\end{equation}
where the momentum~$\vec{p}$ is
%
\begin{equation}
\vec{p} = \vec{V}\,\frac{E_m}{c^2_0}
\label{eq:momentum}
\end{equation}
\citep{Ein05a,Ein05b}.

For an entity in vacuum with $m = 0$, such as a
photon \citep[cf.][]{Ein05c,Lew26,Oku09}, the energy-momentum
relation in Eq.~(\ref{eq:energy}) reduces to
%
\begin{equation}
E_\nu = p_\nu\,c_0 ~.
\label{eq:photon}
\end{equation}

\section{Electrostatic impact model and dipoles}
\label{s.quadrupoles}

In analogy to Eq.~(\ref{eq:photon}), we assumed for hypothetical massless
entities (named ``dipoles'')
%
\begin{equation}
E_{\rm D} = |\vec{p}_{\rm D}|\,c_0 = p_{\rm D}\,c_0~,
\label{eq:dipole}
\end{equation}
where $\vec{p}_{\rm D}$ is the momentum vector of the dipoles,
and constructed an electrostatic impact model \citep[][Paper~2]{Wiletal14}.
The interaction rates of dipoles with bodies~A and B
%
\begin{equation}
\frac{\Del N_{Q,q}}{\Del t} = \frac{\Del N_{q,Q}}{\Del t}
\label{eq:rate}
\end{equation}
(the same for both bodies even for $|Q| \ne |q|$)
required to emulate Coulomb's law in the static case and Newton's
third law can be obtained from Eqs.~(31) and (32) of Paper~2:
%
\begin{eqnarray}
\left|\frac{\Del\vec{P}_{\rm E}(r)}{\Del t}\right| =
p_{\rm D}\,\frac{\Del N_{Q,q}(r)}{\Del t} = \nonumber \\
p_{\rm D}\,\frac{\eta_{\rm E}\,\kappa_{\rm E}}{c_0}\,\frac{|Q|\,|q|}
{4\,\pi\,r^2}  ~ ,
\label{eq:imbalance}
\end{eqnarray}
where $r$ is the separation distance between both bodies and
$|\Del\vec{P}_{\rm E}/\Del t|$ is the norm of the momentum change rate
for~A and B leading together with
%
\begin{equation}
p_{\rm D}\,\eta_{\rm E}\,\kappa_{\rm E} = c_0/\varepsilon_0
\label{eq:relation}
\end{equation}
to an attractive or repulsive electrostatic force of
%
\begin{equation}
F_{\rm E}(r) = \mp |q|\,\frac{|Q|}{4\,\pi\,\varepsilon_0\,r^2} ~.
\label{eq:Coulomb}
\end{equation}
The quantities $\eta_{\rm E}$ and $\kappa_{\rm E}$
are the electrostatic emission and absorption coefficients with the following
definitions:
The absorption coefficient is the dipole absorption rate of a
charge from the background
%
\begin{equation}
\frac{\Del N_Q}{\Del t} =
\kappa_{\rm E}\,\rho_{\rm E}\,|Q| = \eta_{\rm E}\,|Q| ~ ,
\label{eq:ab_em}
\end{equation}
where $\rho_{\rm E} = \Del N_{\rm E}/\Del V$ is the spatial background
number density of dipoles in the volume element $\Del V$, and
$\eta_{\rm E} = \kappa_{\rm E}\,\rho_{\rm E}$ is the emission
coefficient leading to the same emission rate $\Del N_Q/\Del t$.

\section{The potential energy}
\label{s.pot_en}

We may now ask the question, whether the electrostatic impact
model can provide an answer to the ''mysterious'' potential energy problem
in a closed system, where dipoles are interacting with two charged bodies.
The number of dipoles travelling at any instant of time from one charge to
the other can be calculated from the interaction
rate in Eq.~(\ref{eq:imbalance}) multiplied by the travel
time~$\Del t = r/c_0$.
%
\begin{equation}
\Del N_{Q,q}(r) =
\frac{\eta_{\rm E}\,\kappa_{\rm E}}{c_0^2}\,\frac{|Q|\,|q|}{4\,\pi\,r}  ~.
\label{eq:number}
\end{equation}
The same number of dipoles is moving in the opposite direction.
The energy of the dipoles interacting with the corresponding charge then is
%
\begin{eqnarray}
\Del E_{\rm E}(r) = \Del N_{Q,q}(r)\,p_{\rm D}\,c_0 = \nonumber \\
\frac{p_{\rm D}\,\eta_{\rm E}\,\kappa_{\rm E}}{c_0}\,\frac{|Q|\,|q|}
{4\,\pi\,r} = \frac{|Q|\,|q|}{4\,\pi\,\varepsilon_0\,r}  ~.
\label{eq:pot_energy}
\end{eqnarray}
The last term shows\,--\,with reference to
Eqs.~(\ref{eq:potential}) and (\ref{eq:pot_en})\,--\,that the
energy~$\Del E_{\rm E}$ equals the absolute value of the electrostatic
potential energy of body~B
%
\begin{equation}
U_{\rm B}(r) = \phi_{\rm A}(r)\,q = \frac{|Q|\,q}{4\,\pi\,\varepsilon_0\,r} ~,
\label{eq:pot_en}
\end{equation}
at a distance~$r$ from body~A. The symmetry in $Q$ and $q$ implies that
the potential energy of body~A at a distance~$r$ from body~B is the
same. To simplify the following arguments, we will now assume that body~A
has a mass~$m_{\rm A}$ much larger than $m_{\rm B}$ of body~B and can be
considered to be at
rest in an inertial system.
We then calculate the difference of the potential energies for a displacement
of~B from $r$ to $r + \Del r$ as well as the difference of the energies of
the interacting dipoles and get
%
\begin{eqnarray}
U_{\rm B}(r) - U_{\rm B}(r + \Del r) = \nonumber \\
\frac{|Q|\,q}{4\,\pi\,\varepsilon_0}\,\left(\frac{1}
{r} - \frac{1}{r + \Del r}\right) \approx
\frac{|Q|\,q}{4\,\pi\,\varepsilon_0}\,\frac{\Del r}{r^2}
\label{eq:diff_pot}
\end{eqnarray}
and
%
\begin{eqnarray}
\Del E_{\rm E}(r) - \Del E_{\rm E}(r + \Del r) = \nonumber \\
\{\Del N_{Q,q}(r) - \Del N_{Q,q}(r + \Del r)\}\,p_{\rm D}\,c_0 = \nonumber \\
\frac{|Q|\,|q|}{4\,\pi\,\varepsilon_0}\,\left(\frac{1}
{r} - \frac{1}{r + \Del r}\right) \approx
\frac{|Q|\,|q|}{4\,\pi\,\varepsilon_0}\,\frac{\Del r}{r^2} ~,
\label{eq:diff_def}
\end{eqnarray}
where the approximations are valid for $|\Del r| \ll r$.
For $+|q|$ and $\Del r > 0$, Eqs.~(\ref{eq:diff_pot}) and (\ref{eq:diff_def})
correspond to the case of repulsion in Fig.~\ref{fig:plus}. If, on the other
hand, the charge of body~B in Fig.~\ref{fig:minus} is~$-|q|$ and
$\Del r < 0$, the equations describe attraction between the bodies.
In both cases, the result of Eq.~(\ref{eq:diff_pot}) is positive,
i.e. $U_{\rm B}(r) > U_{\rm B}(r + \Del r)$. The difference of the potential
energies can be transformed into kinetic energy with respect to the inertial
system defined.

The last term of Eq.~(\ref{eq:diff_def}) gives the variation of the energy
$\Del E_{\rm E}$ between $r$ and $r + \Del r$. It is positive for repulsion
with $\Del r > 0$ and equal to the result of Eq.~(\ref{eq:diff_pot}).
The source of the potential energy for this process (shown in
Fig.~\ref{fig:plus}) thus is the difference of the number of interacting
dipoles on their way to body~B and the corresponding difference in energy.

In the case of attraction, the result of Eq.~(\ref{eq:diff_def}) is negative,
whereas the difference of the energies in Eq.~(\ref{eq:diff_pot}) was
positive. This can be understood by considering that the number of indirect
interactions in Fig.~\ref{fig:minus} increases and the excess direct
interactions from the background are needed to provide the negative
force~$F_{\rm E}(r)$ in Eq.~(\ref{eq:Coulomb}), which is, however,
controlled by the number of indirect interactions.

A question remains concerning the dipoles travelling to body~A.
Eqs.~(\ref{eq:number}) and (\ref{eq:pot_energy}) are symmetric in $q$ and $Q$
and, therefore, the difference in their number with a distance variation of
$\Del r$ must be the same as that of the dipoles on their way to~B, i.e.,
%
\begin{eqnarray}
\Del N_{Q,q}(r) - \Del N_{Q,q}(r + \Del r) =\nonumber \\
\Del N_{q,Q}(r) - \Del N_{q,Q}(r + \Del r) ~.
\label{eq:solution}
\end{eqnarray}
What happens to the corresponding difference in energy, since body~A in our
approximation is basically at rest? The answer is that the change in
potential energy of body~A at the new relative position with respect to~B
is given by the absolute value of the results of Eq.~(\ref{eq:diff_def})
and thus accounts for the energy difference of Eq.~(\ref{eq:diff_pot}).


\section{Conclusion}
\label{s.concl}

In the framework of a recently proposed electrostatic impact model in Paper~2,
the physical processes related to the variation of the
electrostatic potential energy of two charged bodies have been described and
the ``source region'' of the potential energy in such a system
could be identified. In a configuration with repulsion, the potential energy
is directly related to
the energy of the interacting dipoles on their way from body~A to body~B.
For attraction, the negative force stems from the excess direct
dipole interactions from the background distribution\,--\,in analogy to the
gravitational attraction in Paper~1.


\end{document}